\documentclass{amsart}

\usepackage[round]{natbib}
\usepackage{graphicx}
\usepackage{url}
\usepackage{enumerate}
\usepackage{palatino}
\usepackage{todonotes}
\usepackage{ulem}
\usepackage{subcaption}
\usepackage{lineno}

\newcommand{\mint}{t_{\min}}
\newcommand{\maxt}{t_{\max}}

\newcommand{\arxiv}[1]{#1}
\newcommand{\notarxiv}[1]{}
\newcommand{\eat}[1]{}

\newcommand{\sts}{\textsf{sts}}

\newcommand{\beginsupplement}{%
        \setcounter{table}{0}
        \renewcommand{\thetable}{S\arabic{table}}%
        \setcounter{figure}{0}
        \renewcommand{\thefigure}{S\arabic{figure}}%
     }

\newcommand{\FIGparams}{\
\begin{figure}
\begin{center}
  \arxiv{\includegraphics[width=10.5cm]{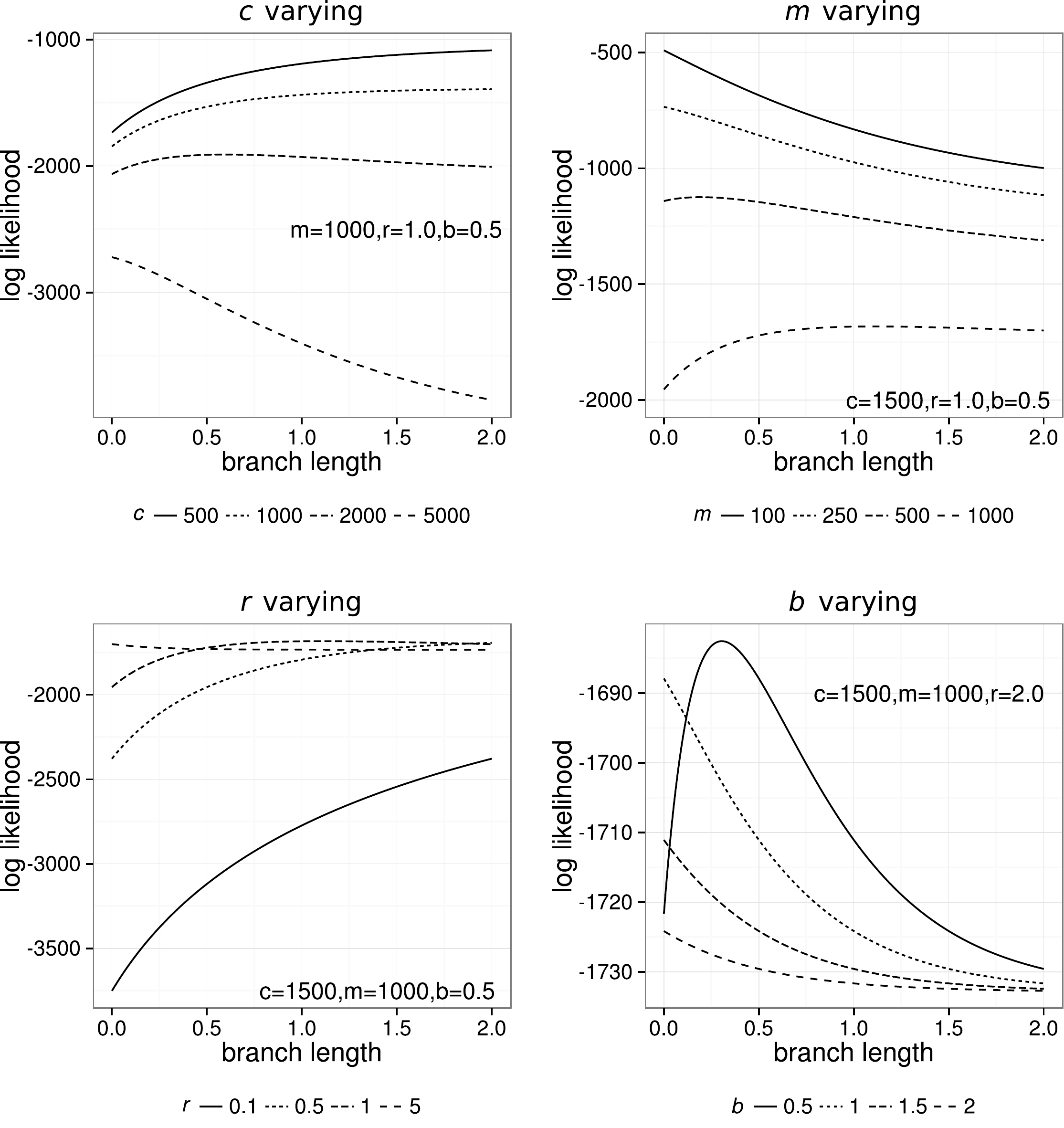}}
\end{center}
\caption{\
  How each of the four parameters changes the shape of our surrogate function $f$ defined in \eqref{eq:surrogate}.
}
\label{FIGparams}
\end{figure}
}

\newcommand{\FIGrelErr}{\
\begin{figure}
\begin{center}
  \arxiv{\includegraphics[width=6in]{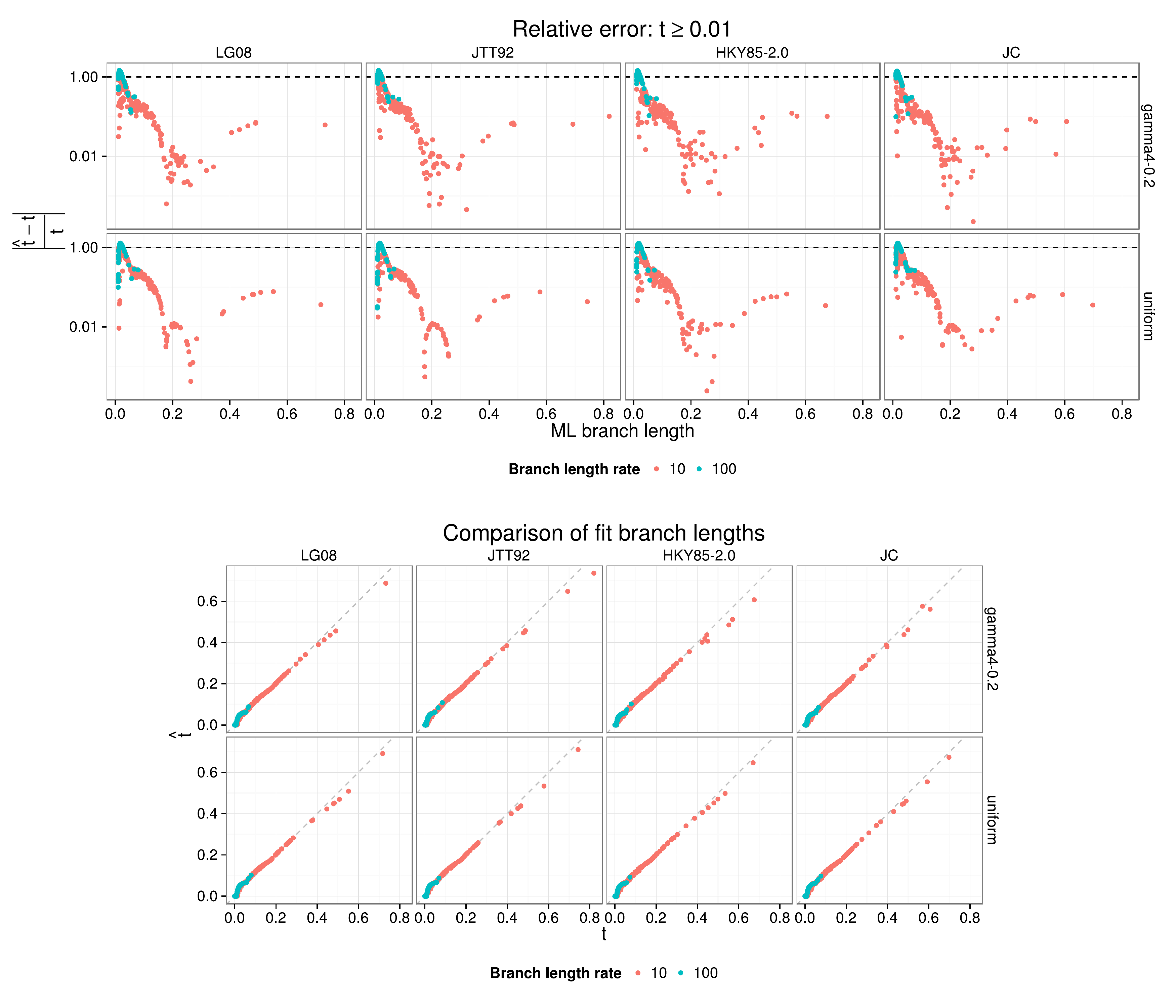}}
\end{center}
\caption{\
  Relative error by branch length.  \emph{To consider -- none / one / both?}
}
\label{FIGrelErr}
\end{figure}
}

\newcommand{\FIGregimes}{\
\begin{figure}
\begin{center}
  \arxiv{\includegraphics[width=7cm]{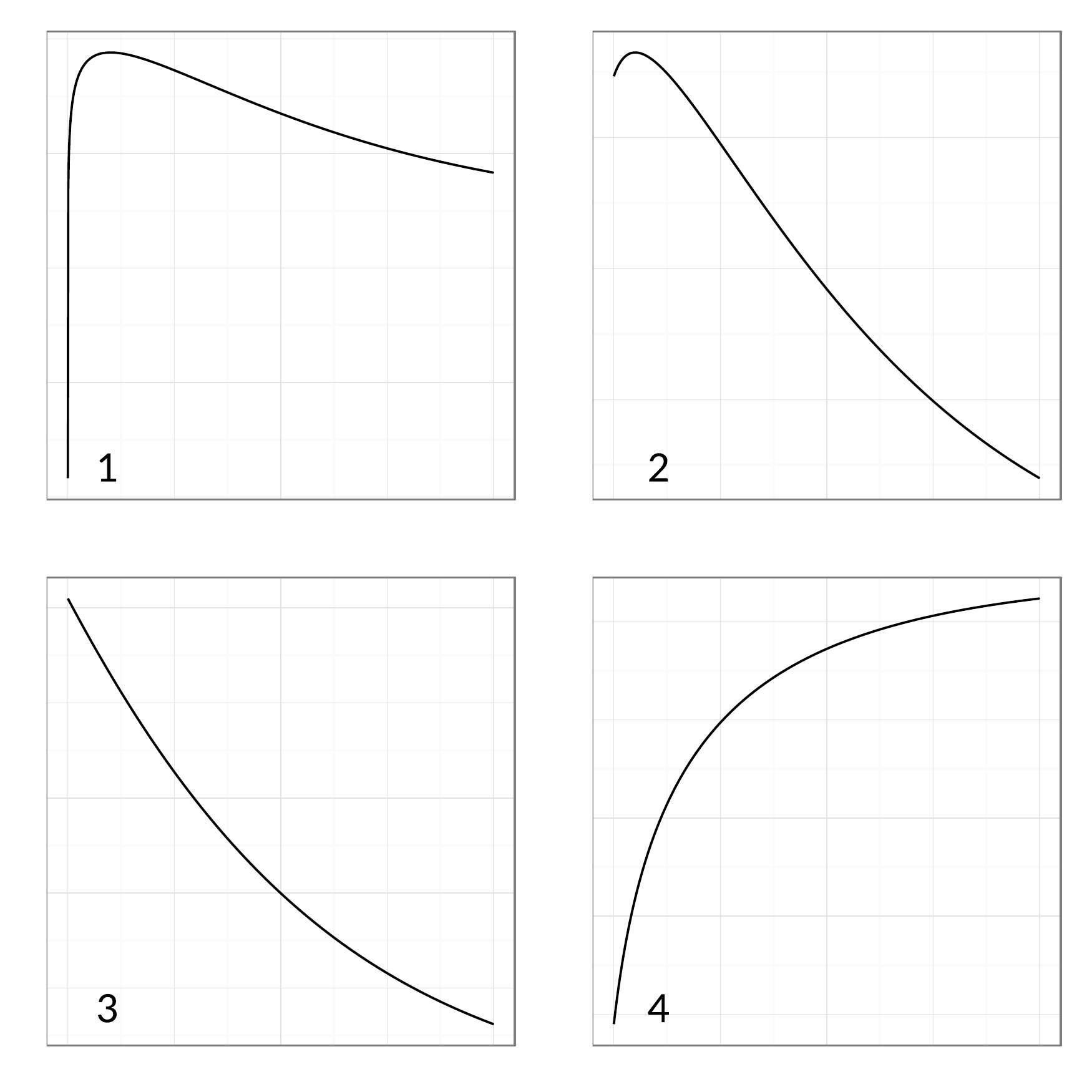}}
\end{center}
\caption{\
  The various regimes of a likelihood function for the BSM parameterized by branch length.
}
\label{FIGregimes}
\end{figure}
}

\newcommand{\FIGkl}{\
\begin{figure}
\begin{center}
\begin{subfigure}{.50\textwidth}
  \includegraphics[width=\linewidth]{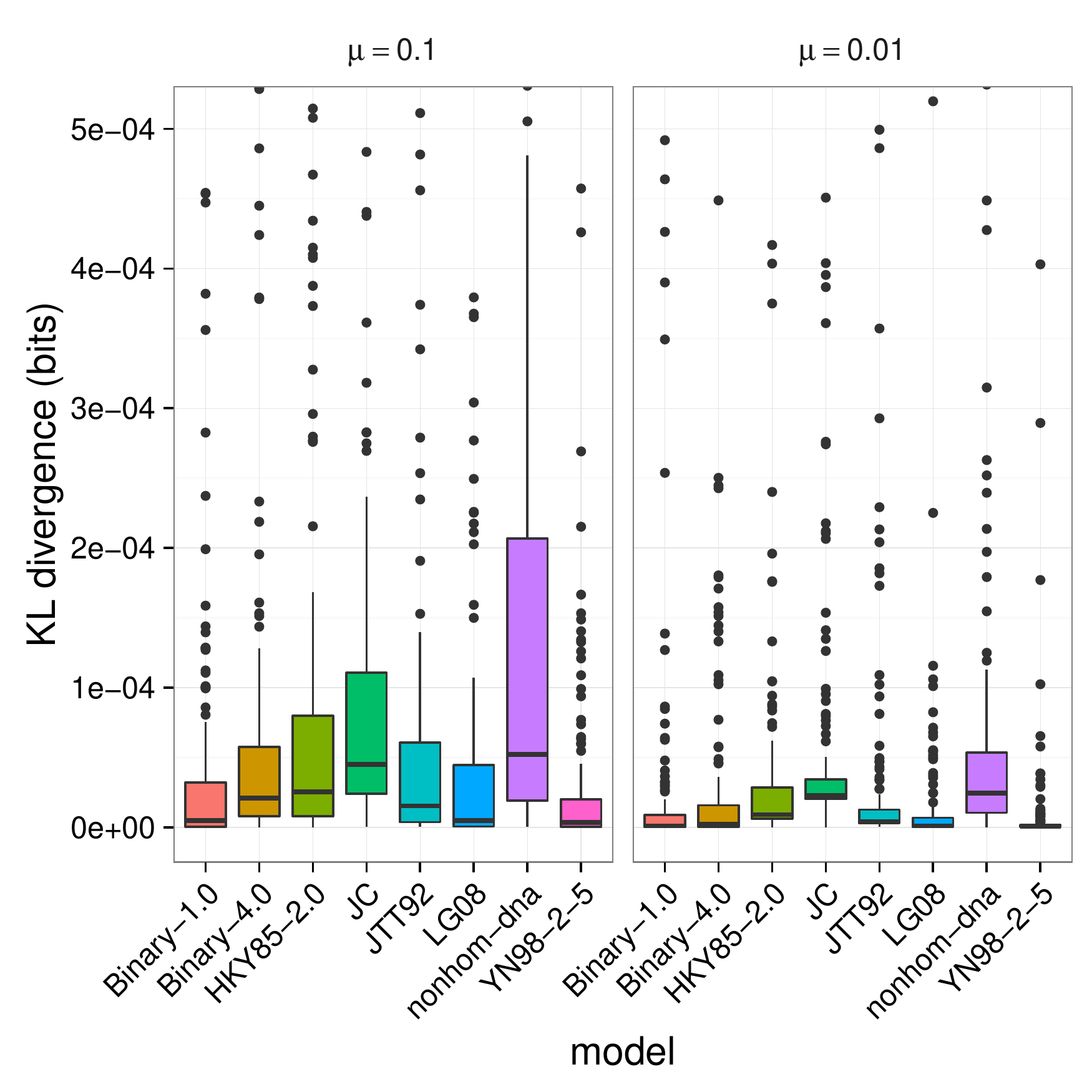}
  \caption{Uniform}
\end{subfigure}%
\begin{subfigure}{.50\textwidth}
  \includegraphics[width=\linewidth]{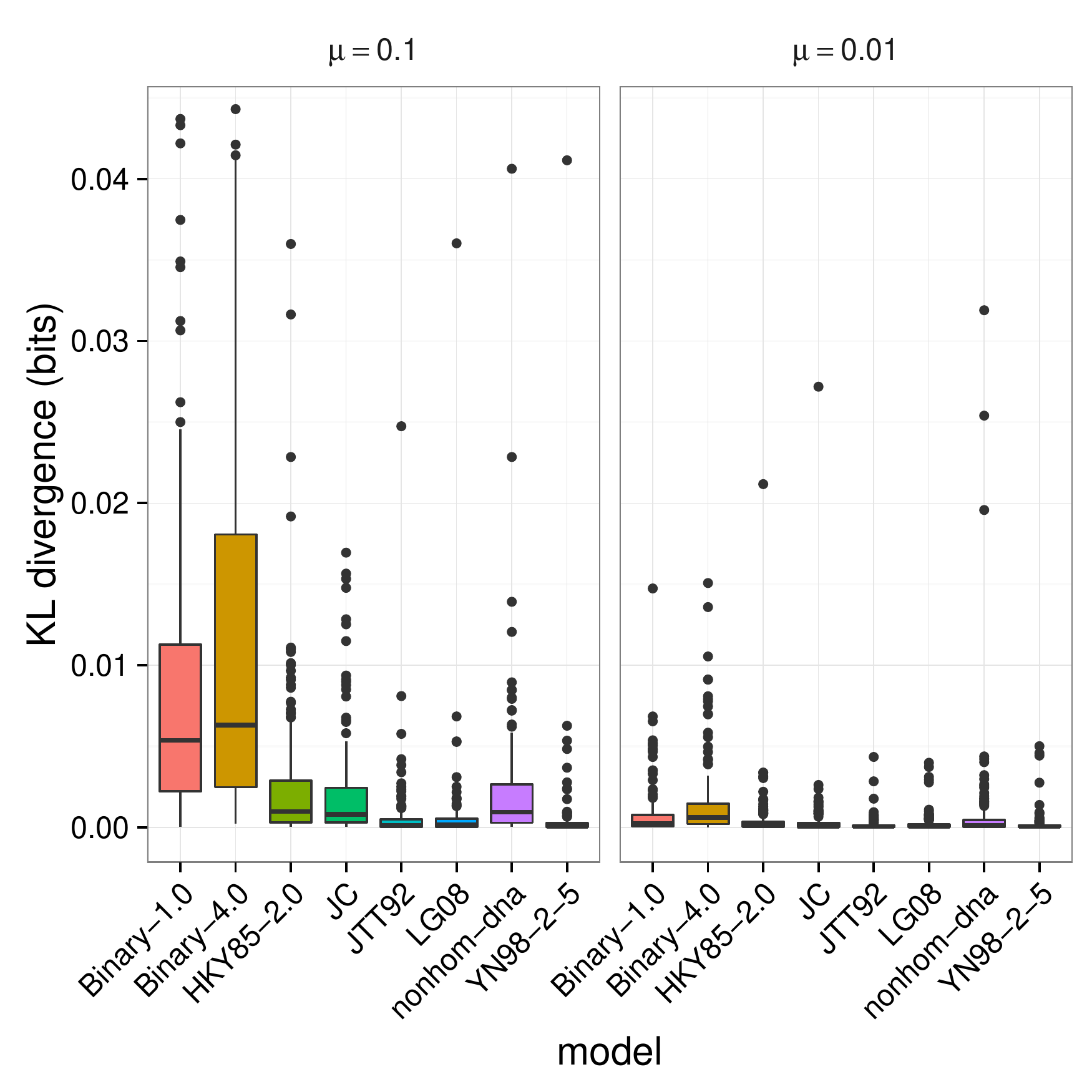}
  \caption{Discretized Gamma}
\end{subfigure}
\end{center}
\caption{\
  Estimated Kullback-Leibler divergence from the original likelihood function to the surrogate function.
  Simulations done using (a) uniform rates across sites and (b) discretized Gamma distributed rates across sites (4 categories, $\alpha=0.2$).
  Branch lengths are either drawn from an exponential with mean either $\mu=0.1$ or $\mu=0.01$.
  See Table~\ref{TABLEmodels} for a list of model name abbreviations.
  Some outlier points excluded for clarity (Table~\ref{TABLEkloutliers}).
}
\label{FIGkl}
\end{figure}
}

\newcommand{\FIGacceptance}{\
\begin{figure}
\begin{center}
  \arxiv{\includegraphics[width=12cm]{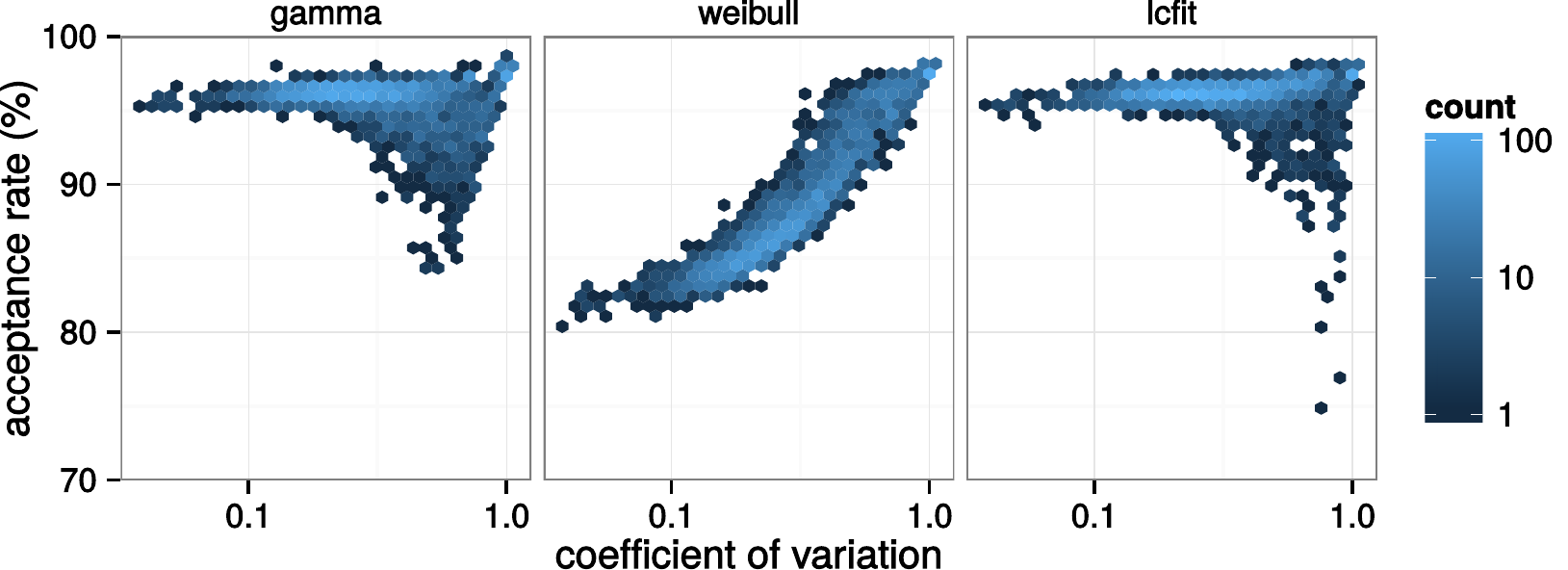}}
\end{center}
\caption{\
  Expected acceptance rate for maximum-likelihood fits of gamma, Weibull, and lcfit distributions versus coefficient of variation of sampled single-branch-length posterior distributions for 12 datasets tested by \cite{Aberer2016-dr}.
  Fit parameters for the gamma and Weibull distributions were obtained directly from data provided by \cite{Aberer2016-dr}; those results reproduced here for comparison to lcfit.
}
\label{FIGacceptance}
\end{figure}
}

\newcommand{\TABLEmodels}{\
\begin{table}[ht]
  \begin{tabular}{llp{3.8cm}p{3.8cm}}
Name        & Data Type  & Parameters                                               & Reference \\
\hline
Binary-1.0     & binary     & $\kappa = 1$                   & see caption         \\
Binary-4.0     & binary     & $\kappa = 4$                   & see caption         \\
JC             & DNA        &                                                          & \citep{JC69}         \\
HKY85          & DNA        & $\kappa = 2.0$, equal base freqs                   & \citep{Hasegawa1985-uj}         \\
JTT92          & amino acid &                                                          & \citep{Jones1992-qv}          \\
LG08           & amino acid &                                                          & \citep{Le2008-cm}         \\
YN98           & codon      & $\kappa = 2.0$, $\omega = 5.0$                           & \citep{Yang1998-ip}         \\
Nonhomogeneous & DNA        & 7~edges with T92 model, 6~edges with TN93 model, 5~edges with GTR & \cite{Tamura1992-sp,Tamura1993-nw,Tavare1986-nb}
\end{tabular}
\caption{\
The models used in Fig.~\ref{FIGkl}.
The binary model is parametrized as in the Bio++ documentation, such that a binary model with parameter $\kappa$ has stationary distribution $(1/(\kappa+1), \kappa/(\kappa+1))$.
}
\label{TABLEmodels}
\end{table}}

\newcommand{\TABLEkloutliers}{\
\begin{table}[ht]
\centering
\begin{tabular}{rllrrrrr}
  \hline
 & Rate Distribution & Branch Length Mean & Count & Plot Threshold & Mean & Median & Maximum \\
  \hline
1 & gamma4-0.2 & mu == 0.1 &    23 & 0.0415 & 0.0673 & 0.0577 & 0.2485 \\
  2 & gamma4-0.2 & mu == 0.01 &     5 & 0.0415 & 0.1011 & 0.0898 & 0.1235 \\
  3 & uniform & mu == 0.1 &   123 & 0.0005 & 0.0073 & 0.0013 & 0.2020 \\
  4 & uniform & mu == 0.01 &    65 & 0.0005 & 0.0629 & 0.0016 & 3.6035 \\
   \hline
\end{tabular}
\caption{\
Outlier points excluded from Fig.~\ref{FIGkl}.
The counts are out of $1370$ edges evaluated for each rate distribution/branch length mean combination.
}
\label{TABLEkloutliers}
\end{table}}

\title{A surrogate function for one-dimensional phylogenetic likelihoods}
\author[Claywell]{Brian C. Claywell}
\author[Dinh]{Vu C. Dinh}
\author[McCoy]{Connor O. McCoy}
\author[Matsen]{Frederick A. Matsen IV}
\date{\today}

\begin{document}

\begin{abstract}
Phylogenetics has seen an steady increase in substitution model complexity, which requires increasing amounts of computational power to compute likelihoods.
This model complexity motivates strategies to approximate the likelihood functions for branch length optimization and Bayesian sampling.
In this paper, we develop an approximation to the one-dimensional likelihood function as parametrized by a single branch length.
This new method uses a four-parameter surrogate function abstracted from the simplest phylogenetic likelihood function, the binary symmetric model.
We show that it offers a surrogate that can be fit over a variety of branch lengths, that it is applicable to a wide variety of models and trees, and that it can be used effectively as a proposal mechanism for Bayesian sampling.
The method is implemented as a stand-alone open-source C library for calling from phylogenetics algorithms; it has proven essential for good performance of our online phylogenetic algorithm \sts.
\end{abstract}

\maketitle

\section{Introduction}
The increasing availability of large molecular sequence data sets poses a challenge for current phylogenetic algorithms.
At the same time, phylogenetic substitution models are becoming more realistic and consequently, more complex \citep{lartillot2004bayesian,zoller2012improving,Groussin2013-jw,Wang2014-ij}.
The combination of a large and increasing amount of phylogenetic likelihood calculation along with increasing complexity of models motivates research into useful approximations to the phylogenetic likelihood function.

One simple opportunity for efficiency improvement is in optimization of, or sampling from, the likelihood function as parametrized by a single branch length while fixing other parameters.
In this case the likelihood function is simply a function that takes a non-negative real input and gives out another real number.
One common approach for numerical maximization of such functions $\ell$ is to sample an $\ell$ at a number of points, fit a simple curve to those points, and then use the fit as an approximation to $\ell$.
We will call $\ell$ the \emph{original function} and the fitted function $f$ the \emph{surrogate function}.
Such an approach is useful if the original function is expensive to evaluate, but the surrogate function can be quickly fit to the sample points and evaluated.
It is already being used implicitly in phylogenetics by inference programs that use Brent's method \citep{brent1973algorithms} for likelihood maximization, a method which effectively uses linear interpolation via the secant method.
Recent work by \cite{Aberer2016-dr} shows that proposals built using common probability distribution functions (PDFs) as surrogates, in particular the $\Gamma$ distribution, can have high acceptance rates.
Bayesian statistics in general has benefited from the use of likelihood function approximations, such as for variational analysis \citep{Wainwright2008-vd}.

Although known functions can provide useful surrogates in phylogenetics, one might desire a class of surrogate functions that is specialized to the task.
Indeed, phylogenetic likelihood functions parameterized by a single branch length have special characteristics: they asymptote as the branch length becomes long, and sometimes achieve infinite slope as the branch length becomes short.
Neither of these features can be true for any polynomial, nor are they true for PDFs of common distribution functions.

In this paper, we show that a slight generalization of the likelihood function for the binary symmetric model (BSM) on a two-taxon tree can serve as a useful surrogate function for likelihood functions parameterized by branch lengths.
We call this surrogate the \emph{lcfit} function, short for ``likelihood curve fit.''
With only four parameters, it can be easily and efficiently fit in a least-squares sense with standard algorithms; even more robust fitting can be achieved using the ML branch length and corresponding second derivative.
We show via experiments with simulated and real data that it is readily fit and does a good job of approximating even complex models, making it a useful tool when those models are expensive to evaluate.
Our code to use lcfit is available as an open-source C library.

\section{Results}
\subsection{Surrogate formula and fitting}

\arxiv{\FIGparams}

The lcfit surrogate function $f$ evaluated at branch length $t$ is
\begin{equation}
f(c,m,r,b;t) = c \log[(1+e^{-r(t+b)})/2] + m \log[(1-e^{-r(t+b)})/2]
\label{eq:surrogate}
\end{equation}
for any positive values of the lcfit coefficients $c,m,r$, and non-negative $b$.
It can be considered as an abstract surrogate function that takes a set of shapes resembling those of phylogenetic likelihood curves (Fig.~\ref{FIGparams}).
However, when $b$ is zero this function is the log likelihood function for the binary symmetric model (BSM; see, e.g., Semple and Steel, 2003\nocite{semple2003phylogenetics}) where $c$ is the number of constant sites, $m$ is the number of substituted sites, and $r$ is the substitution rate.
The inclusion of the $b$ term simply serves to truncate the likelihood function on the left, which is helpful in fitting likelihood functions for trees with more than two taxa.
Indeed, without truncation the limit of $f$ as branch lengths go to $0$ is always negative infinity; this does not typically make for a good fit to likelihood functions parameterized by branches of non-trivial phylogenetic trees.
As the branch length becomes long, $f$ approaches an asymptote of $-(c+m) \log(2)$.

We will assume that $r > 0$ and $b \geq 0$, so that $e^{-r(t+b)}$ as a function of non-negative $t$ goes from some positive value down to zero.
The maximum of the log likelihood function for this setting is
\begin{equation}
t_0 = -b + \log[(c+m)/(c-m)] / r.
\label{eq:BSMmlt}
\end{equation}
This has a finite real solution exactly when $c>m$.
In the BSM interpretation this means that the number of constant sites strictly exceeds the number of substituted sites.
Other characteristics of the lcfit function $f$ are easily derived, such as the second derivative at the maximum, and the inflection point when it exists (see Supplement for formulas and derivations).
Using such formulas we have found it useful in some cases to re-parameterize $f$ in terms of the original $c$, $m$, $f$'s maximum $t_0$, and the second derivative at this maximum value $f''(t_0)$.

Briefly, our fitting methods combine two strategies to fit the parameters of the lcfit function (details provided in the Supplement).
Both use least-squares fitting of sampled branch lengths and their likelihoods.
The first strategy (lcfit2) applies when the maximum likelihood branch length is positive, and uses the second derivative at this branch length to eliminate two parameters so that only two parameters need to be fit.
The second strategy (lcfit4) simply fits the lcfit parameters using least-squares directly.

We can simply multiply an lcfit curve by a branch length prior to get an approximate (unnormalized) PDF.
For sampling from this lcfit PDF we have used a simple rejection sampling strategy with an exponential proposal distribution.
Although this may require many proposals for an acceptance for certain lcfit shapes, individual lcfit evaluations are computationally cheap so we have not found this to be a significant burden in practice.

C library code with unit tests, continuous testing, simulation framework, and documentation is available at \url{https://github.com/matsengrp/lcfit}.

\subsection{Performance}

We obtain slightly better results than \citet{Aberer2016-dr} in terms of acceptance rate for branch length proposals using their benchmarking strategy (Fig.~\ref{FIGacceptance}).
Briefly, we re-used their acceptance rate results for their $\Gamma$ and Weibull proposals and used the same trees and likelihoods to compute the lcfit surrogate function (see Supplementary~Methods for details).
In terms of computational time, both our method and the method of \citet{Aberer2016-dr} require the maximum of the likelihood function to be found, along with the second derivative.
This computational effort dominates the required effort, and thus they are approximately equal in terms of computational cost.
\FIGacceptance

We then performed simulation to explore how well the lcfit surrogate fits a broader range of models.
To do so, we simulated data under a variety of models, and fit lcfit to the resulting likelihood curves under the same models.
We quantified the divergence between the two curves using Kullback-Leibler (KL) divergence.
We found that KL divergence for complex models is similar to KL divergence for data simulated under binary model (Fig.~\ref{FIGkl}).
Surprisingly, we found that lcfit performance by this metric was worse for variants of the binary model (e.g. the non-symmetric binary model or a mixture of rates) than for more complex models.
\FIGkl

\section{Discussion}

In this paper we present lcfit, the first surrogate function specialized to the case of one-dimensional phylogenetic likelihood functions, and how it can be useful.
Our work shares goals with those of \cite{Aberer2016-dr}, however there are several aspects of our framework that make it appealing.
This previous work uses several standard probability distributions as surrogate functions for posteriors.
In particular, they fit normal, lognormal, Weibull, and $\Gamma$ distributions to approximate per-branch posterior distributions in order to obtain efficient proposals.
With the best performing of these distributions (typically $\Gamma$) they obtain high acceptance rates.
However, there are inherent limitations using standard distributions.
For example, the $\Gamma$ and Weibull have two different shapes, depending on if their shape parameter is greater and less than one; when the shape parameter is greater than one, the value at zero is zero, and when it is less than one then the first derivative at zero is negative.
Neither of these need hold for phylogenetic likelihood curves or posteriors.
Indeed, likelihood curves for internal branches are typically nonzero at zero and have a nonzero modes, for example, see Fig.~1c of \cite{Aberer2016-dr}.
The truncated normal can take this shape, but its symmetry makes it a bad choice in this setting.
In addition, lcfit matches real per-branch likelihoods by enabling a nonzero asymptote, whereas the \cite{Aberer2016-dr} surrogates are all zero at infinity.

In addition to theoretical advantages of the lcfit framework, there are several practical advantages.
\cite{Aberer2016-dr} develop a fitting procedure using a linear relationship between the second derivative of the likelihood function and the standard deviation of the posterior density of the branch length.
However, to use this relationship the parameters of this linear relationship must be inferred.
Because it is inefficient to infer these parameters on the fly, \cite{Aberer2016-dr} use consensus values and a somewhat complex tuning procedure, whereas in most cases we simply fit two coefficients using standard least-squares methods.
We also note that lcfit is implemented as a stand-alone library for incorporation into other software, whereas the independence sampler of \cite{Aberer2016-dr} is baked into ExaBayes \citep{Aberer2014-gy}.

We have found lcfit to be essential for an efficient implementation \citep{sts} of Online Phylogenetic Sequential Monte Carlo \citep{Dinh2016-vw}; this work also points the way to needed extensions.
Here we have focused on approximating phylogenetic likelihood as a function of a single branch length at a time, but one could similarly concoct surrogate functions for other low-dimensional settings.
For example, one could maximize three branches around an internal node by using a surrogate function based on the BSM likelihood function for a three taxon tree, or consider branch length changes and nearest-neighbor interchange moves simultaneously by using a surrogate function based on the BSM likelihood function for a four taxon tree.

\section{Acknowledgements}
The authors would like to thank Steve Evans, Vladimir Minin, Aaron Darling, Chris Warth, Andr\'e Aberer, Julien Dutheil and Bastien Boussau.
This work was supported by National Science Foundation awards DMS-1223057 and CISE-1564137, and National Institutes of Health grant U54GM111274.
The research of Frederick Matsen was supported in part by a Faculty Scholar grant from the Howard Hughes Medical Institute and the Simons Foundation.

\clearpage

\bibliography{lcfittex}

\begin{thebibliography}{}

\bibitem[Aberer {\em et~al.}(2014)Aberer, Kobert, and
  Stamatakis]{Aberer2014-gy}
Aberer, A.~J., Kobert, K., and Stamatakis, A. 2014.
\newblock {ExaBayes}: massively parallel bayesian tree inference for the
  whole-genome era.
\newblock {\em Mol. Biol. Evol.}, {31}(10): 2553--2556.

\bibitem[Aberer {\em et~al.}(2016)Aberer, Stamatakis, and
  Ronquist]{Aberer2016-dr}
Aberer, A.~J., Stamatakis, A., and Ronquist, F. 2016.
\newblock An efficient independence sampler for updating branches in bayesian
  markov chain monte carlo sampling of phylogenetic trees.
\newblock {\em Syst. Biol.}, {65}(1): 161--176.

\bibitem[Brent(1973)Brent]{brent1973algorithms}
Brent, R. 1973.
\newblock {\em Algorithms for minimization without derivatives\/}.
\newblock Prentice-Hall.

\bibitem[Davis and Polonsky(1964)Davis and Polonsky]{davis1964numerical}
Davis, P.~J. and Polonsky, I. 1964.
\newblock Numerical interpolation, differentiation, and integration.
\newblock In M.~Abramowitz and I.~A. Stegun, editors, {\em Handbook of
  Mathematical Functions: With Formulas, Graphs, and Mathematical Tables\/},
  chapter~25. U.S. Government Printing Office, tenth edition.

\bibitem[Dinh {\em et~al.}(2016)Dinh, Darling, and Matsen]{Dinh2016-vw}
Dinh, V., Darling, A.~E., and Matsen, IV, F.~A. 2016.
\newblock Online bayesian phylogenetic inference: theoretical foundations via
  sequential monte carlo.

\bibitem[Dutheil and Boussau(2008)Dutheil and Boussau]{dutheil2008non}
Dutheil, J. and Boussau, B. 2008.
\newblock Non-homogeneous models of sequence evolution in the {Bio++} suite of
  libraries and programs.
\newblock {\em BMC Evolutionary Biology\/}, {8}(1): 255.

\bibitem[Dutheil {\em et~al.}(2006)Dutheil, Gaillard, Bazin, Gl{\'e}min,
  Ranwez, Galtier, and Belkhir]{dutheil2006biopp}
Dutheil, J., Gaillard, S., Bazin, E., Gl{\'e}min, S., Ranwez, V., Galtier, N.,
  and Belkhir, K. 2006.
\newblock {Bio++: a set of C++ libraries for sequence analysis, phylogenetics,
  molecular evolution and population genetics}.
\newblock {\em BMC bioinformatics\/}, {7}(1): 188.

\bibitem[Fourment {\em et~al.}(2017)Fourment, Claywell, Dinh, McCoy, Matsen~IV,
  and Darling]{sts}
Fourment, M., Claywell, B.~C., Dinh, V.~C., McCoy, C.~O., Matsen~IV, F.~A., and
  Darling, A.~E. 2017.
\newblock {Effective online Bayesian phylogenetics via sequential Monte Carlo
  with guided proposals}.
\newblock {\em In preparation\/}.

\bibitem[Galassi and Gough(2003)Galassi and Gough]{galassi2003gnu}
Galassi, M. and Gough, B. 2003.
\newblock {\em {GNU} Scientific Library: Reference Manual : Edition 1.6\/}.
\newblock Network Theory.

\bibitem[Groussin {\em et~al.}(2013)Groussin, Boussau, and
  Gouy]{Groussin2013-jw}
Groussin, M., Boussau, B., and Gouy, M. 2013.
\newblock A {Branch-Heterogeneous} model of protein evolution for efficient
  inference of ancestral sequences.
\newblock {\em Syst. Biol.}

\bibitem[Hasegawa {\em et~al.}(1985)Hasegawa, Kishino, and
  Yano]{Hasegawa1985-uj}
Hasegawa, M., Kishino, H., and Yano, T. 1985.
\newblock Dating of the human-ape splitting by a molecular clock of
  mitochondrial {DNA}.
\newblock {\em J. Mol. Evol.}, {22}(2): 160--174.

\bibitem[Johnson(2014)Johnson]{Johnson2014-ww}
Johnson, S.~G. 2014.
\newblock The {NLopt} nonlinear-optimization package.
\newblock \url{http://ab-initio.mit.edu/nlopt}.

\bibitem[Jones {\em et~al.}(1992)Jones, Taylor, and Thornton]{Jones1992-qv}
Jones, D.~T., Taylor, W.~R., and Thornton, J.~M. 1992.
\newblock The rapid generation of mutation data matrices from protein
  sequences.
\newblock {\em Comput. Appl. Biosci.}, {8}(3): 275--282.

\bibitem[Jukes and Cantor(1969)Jukes and Cantor]{JC69}
Jukes, T.~H. and Cantor, C.~R. 1969.
\newblock Evolution of protein molecules.
\newblock In H.~N. Munro, editor, {\em Mammalian protein metabolism\/}, pages
  21--132. Academic Press, New York.

\bibitem[Kraft(1994)Kraft]{kraft1994slsqp}
Kraft, D. 1994.
\newblock Algorithm 733: {TOMP}--{Fortran} modules for optimal control
  calculations.
\newblock {\em ACM Trans. Math. Softw.}, {20}(3): 262--281.

\bibitem[Lartillot and Philippe(2004)Lartillot and
  Philippe]{lartillot2004bayesian}
Lartillot, N. and Philippe, H. 2004.
\newblock A {Bayesian} mixture model for across-site heterogeneities in the
  amino-acid replacement process.
\newblock {\em Molecular biology and evolution\/}, {21}(6): 1095--1109.

\bibitem[Le and Gascuel(2008)Le and Gascuel]{Le2008-cm}
Le, S.~Q. and Gascuel, O. 2008.
\newblock An improved general amino acid replacement matrix.
\newblock {\em Mol. Biol. Evol.}, {25}(7): 1307--1320.

\bibitem[Levenberg(1944)Levenberg]{levenberg1944method}
Levenberg, K. 1944.
\newblock A method for the solution of certain non-linear problems in least
  squares.
\newblock {\em Quarterly of Applied Mathematics\/}, {2}: 164--168.

\bibitem[Marquardt(1963)Marquardt]{marquardt1963algorithm}
Marquardt, D. 1963.
\newblock An algorithm for least-squares estimation of nonlinear parameters.
\newblock {\em Journal of the Society for Industrial \& Applied Mathematics\/},
  {11}(2): 431--441.

\bibitem[McCoy {\em et~al.}(2012)McCoy, Gallagher, Hoffman, and
  Matsen]{mccoy2012nestly}
McCoy, C., Gallagher, A., Hoffman, N., and Matsen, F. 2012.
\newblock nestly-- a framework for running software with nested parameter
  choices and aggregating results.
\newblock {\em Bioinformatics\/}.

\bibitem[Paradis {\em et~al.}(2004)Paradis, Claude, and
  Strimmer]{paradis2004ape}
Paradis, E., Claude, J., and Strimmer, K. 2004.
\newblock A{PE}: analyses of phylogenetics and evolution in {R} language.
\newblock {\em Bioinformatics\/}, {20}: 289--290.

\bibitem[Semple and Steel(2003)Semple and Steel]{semple2003phylogenetics}
Semple, C. and Steel, M. 2003.
\newblock {\em Phylogenetics\/}.
\newblock Oxford University Press.

\bibitem[Tamura(1992)Tamura]{Tamura1992-sp}
Tamura, K. 1992.
\newblock Estimation of the number of nucleotide substitutions when there are
  strong transition-transversion and {G+C-content} biases.
\newblock {\em Mol. Biol. Evol.}, {9}(4): 678--687.

\bibitem[Tamura and Nei(1993)Tamura and Nei]{Tamura1993-nw}
Tamura, K. and Nei, M. 1993.
\newblock Estimation of the number of nucleotide substitutions in the control
  region of mitochondrial {DNA} in humans and chimpanzees.
\newblock {\em Mol. Biol. Evol.}, {10}(3): 512--526.

\bibitem[Tavar{\'e}(1986)Tavar{\'e}]{Tavare1986-nb}
Tavar{\'e}, S. 1986.
\newblock Some probabilistic and statistical problems in the analysis of {DNA}
  sequences.
\newblock {\em Lectures on mathematics in the life sciences\/}.

\bibitem[Wainwright and Jordan(2008)Wainwright and Jordan]{Wainwright2008-vd}
Wainwright, M.~J. and Jordan, M.~I. 2008.
\newblock Graphical models, exponential families, and variational inference.
\newblock {\em Foundations and Trends\textregistered{} in Machine Learning\/},
  {1}(1--2): 1--305.

\bibitem[Wang {\em et~al.}(2014)Wang, Susko, and Roger]{Wang2014-ij}
Wang, H.-C., Susko, E., and Roger, A.~J. 2014.
\newblock An amino acid substitution-selection model adjusts residue fitness to
  improve phylogenetic estimation.
\newblock {\em Mol. Biol. Evol.}

\bibitem[Yang and Nielsen(1998)Yang and Nielsen]{Yang1998-ip}
Yang, Z. and Nielsen, R. 1998.
\newblock Synonymous and nonsynonymous rate variation in nuclear genes of
  mammals.
\newblock {\em J. Mol. Evol.}, {46}(4): 409--418.

\bibitem[Zoller and Schneider(2012)Zoller and Schneider]{zoller2012improving}
Zoller, S. and Schneider, A. 2012.
\newblock Improving phylogenetic inference with a semiempirical amino acid
  substitution model.
\newblock {\em Molecular Biology and Evolution\/}.

\end{thebibliography}
\bibliographystyle{natbib}

\clearpage

\beginsupplement

\section{Supplementary Methods}

\subsection{Parameter regimes for the surrogate function}
For brevity, we define
\[
\theta = \exp(r(t+b)).
\]
We will assume that $r > 0$ and $b \geq 0$, so that $\theta$ as a function of non-negative $t$ goes from some value greater than or equal to 1 up to infinity.
Also note that $d \theta/dt = r \theta$ and $d \theta^{-1}/dt = - r \theta^{-1}$.

The surrogate function is defined as
\begin{align*}
f(c,m,r,b;t) & = c \log((1+\theta^{-1})/2) + m \log((1-\theta^{-1})/2) \\
& = c \log(1+\theta^{-1}) + m \log(1-\theta^{-1}) - (c+m) \log 2.
\end{align*}
As $t$ goes to infinity, this has limit $- (c + m) \log 2$.

Taking the derivative,
\begin{align*}
d f/dt & = -c r \theta^{-1}/(1+\theta^{-1}) + m r \theta^{-1}/(1-\theta^{-1}) \\
& = \frac{-c r}{\theta+1} + \frac{m r}{\theta-1} \\
& = r (-c \theta + c + m \theta + m)/(\theta^2-1) \\
& = r ((m-c) \theta + m+c)/(\theta^2-1).
\end{align*}
So the first derivative is zero when (using subscript zero to denote maximum) $\theta_0 = (c+m)/(c-m);$ this gives a finite real solution for $t$ when $c>m$.
This is equivalent to
\begin{equation}
t_0 = -b + \log[(c+m)/(c-m)] / r.
\label{eq:firstDerivZero}
\end{equation}

For a more complete characterization of $f$, we also take the second derivative:
$$
\frac{d^2 f}{dt^2} =
r^2 \theta \frac{(c-m) (\theta^2+1) -2 (c+m) \theta}{(\theta^2-1)^2}
$$

This is zero when
\begin{equation*}
\exp(r(t+b)) = \theta = \frac{\left(\sqrt{c} \pm \sqrt{m}\right)^2}{c-m};
\end{equation*}
or
\begin{equation}
t = -b + \frac{1}{r} \log \left(\frac{\left(\sqrt{c} \pm \sqrt{m}\right)^2}{c-m} \right);
\label{eq:secondDerivZero}
\end{equation}

We also note that $c > m$ implies $c-m > (\sqrt{c} - \sqrt{m})^2$, meaning that there can never be two positive solutions.
With this we distinguish between four regimes:
\begin{enumerate}[1.]
    \item one negative and one positive root of \eqref{eq:secondDerivZero}, $f(t)$ diverges at $t=0$: $b = 0$, $c > m$ and $\exp(b r) \leq (\sqrt{c} + \sqrt{m})^2)/(c-m)$
    \item one negative and one positive root of \eqref{eq:secondDerivZero}, $f(t)$ finite for all $t \ge 0$: $b > 0$, $c > m$ and $\exp(b r) \leq (\sqrt{c} + \sqrt{m})^2)/(c-m)$
    \item two negative solutions of \eqref{eq:secondDerivZero}: $c > m$ and $\exp(b r) > (\sqrt{c} + \sqrt{m})^2/(c-m)$
    \item no real solutions of \eqref{eq:secondDerivZero}: $c < m$.
\end{enumerate}

This determines the shape of the likelihood curve up to the sign of the second derivative (Fig.~\ref{FIGregimes}) for positive $t$.
Only in cases (1) and (2) are there inflection points.
Only in cases (1) and (4) is the limit as $t$ goes to zero infinite.
In (3) and (4) the ML $t$ is zero and infinity, respectively.
Assuming a tree with finite branch lengths, note that the probability of having something in (4) goes to zero as sequences become long.

\FIGregimes

\subsection{lcfit2 parameterization}
It can be useful to use an alternative parameterization to \ref{eq:surrogate}.
The ``lcfit2'' parameterization is in terms of $c$, $m$, the branch length $t_0$ giving the maximum value of the surrogate, and the second derivative at $t_0$.
We assume that we are in parameter regime 1 or 2, so $c > m$.

We can re-express everything in terms of the difference from the ML branch length $t_0$ and eliminate $b$.
Let $\tilde t$ be $t - t_0$ and $\tilde \theta = \exp(r(t-t_0))$.
Note that $\theta = \tilde \theta \theta_0$, so we can re-express $f$ in these terms, recalling that $\theta_0 = (c+m)/(c-m)$:
\begin{align*}
f(&c,m,r,t_0; \tilde t) \\
& = c \log \left(1+(\tilde \theta \theta_0)^{-1}\right) + m \log \left(1-(\tilde \theta \theta_0)^{-1}\right) - (c+m) \log 2 \\
& = c \log \left(1+ \frac{c-m}{\tilde \theta (c+m)}\right) + m \log \left(1-\frac{c-m}{\tilde \theta (c+m)}\right) - (c+m) \log 2 \\
& = c \log \left(c+m+ \frac{c-m}{\tilde \theta}\right) + m \log \left(c+m-\frac{c-m}{\tilde \theta}\right) \\
& \qquad \qquad \qquad \qquad - (c+m) \log (c+m) - (c+m) \log 2 \\
& = c \log \left(c+m+ \frac{c-m}{\tilde \theta}\right) + m \log \left(c+m-\frac{c-m}{\tilde \theta}\right) - (c+m) \log (2 (c+m)) \\
\end{align*}
Also recall
\begin{align*}
f'(t) = \frac d{d t} f(c,m,r,b; t) & = \frac{-c r}{\theta+1} + \frac{m r}{\theta-1} \\
f''(t) = \frac {d^2}{d t^2} f(c,m,r,b; t) & = \frac{c r^2 \theta}{(\theta+1)^2} + \frac{-m r^2 \theta}{(\theta-1)^2}.
\end{align*}
At the ML point $t_0$, note
$$
\theta_0 + 1 = \frac{2c}{c-m} \qquad \theta_0 - 1 = \frac{2m}{c-m}
$$
so
$$
\frac{\theta_0}{(\theta_0 + 1)^2} = \frac{(c-m) (c+m)}{4c^2} \qquad \frac{\theta_0}{(\theta_0 - 1)^2} = \frac{(c-m) (c+m)}{4m^2}
$$
and
\begin{align*}
f''(t_0) & = r^2 \left( \frac{(c-m)(c+m)}{4c} - \frac{(c-m)(c+m)}{4m} \right) \\
& = r^2  \frac{(c-m)(c+m)}{4} \left( \frac{1}{c} - \frac{1}{m} \right) \\
& = r^2 \frac{(c-m)(c+m)}{4} \left( \frac{m-c}{cm} \right) \\
& = - r^2 \frac{(c-m)^2 (c+m)}{4cm}.
\end{align*}
So
$$
r = \frac{2}{c-m} \sqrt{\frac{-f''(t_0) c m}{c+m}}.
$$

\subsection{Sampling from the PDF corresponding to the surrogate function}
In the context of a Bayesian Monte Carlo algorithm, we can use the fit likelihood curve to quickly draw proposals from an approximate unnormalized posterior, which is simply the lcfit likelihood function times a prior.
For example, we have found this useful in the context of online-sts.
To draw such proposals, we can first use the procedure detailed above to fit an approximate likelihood curve and then use rejection sampling to draw from the approximate posterior.

Rejection sampling generates samples from an arbitrary distribution $h(x)$ using a proposal distribution $g(x)$ subject only to the constraint that $h(x) \leq c g(x)$ for some constant $c > 0$.
For an exponential prior, let $h(t)$ be the unnormalized posterior on branch lengths
\begin{equation*}
  h(t) = \lambda e^{- \lambda t} F(t)
\end{equation*}
where $F(t) = e^{f(t)}$ is the surrogate likelihood function for some set of fit parameters.
Let $g(t)$ be the PDF of the exponential distribution with rate $\lambda$,
\begin{equation*}
  g(t) = \lambda e^{- \lambda t}.
\end{equation*}
Clearly the ratio $h(t) / g(t) = F(t)$, so we choose $c$ to be the maximum likelihood value
\begin{equation*}
  c = F(t_0)
\end{equation*}
where $t_0$ is the mode of the surrogate function and can be computed directly using \eqref{eq:BSMmlt}.

The procedure for generating a sample from the distribution begins by drawing a branch length $t$ from the exponential distribution with rate $\lambda$ and a value $u$ from the uniform distribution over $(0, 1]$.
If
\begin{equation*}
  u \leq \frac{h(t)}{c g(t)} = \frac{F(t)}{F(t_0)}
\end{equation*}
the sample is accepted; otherwise, the sample is rejected and the procedure is repeated.
We note that eliminating the prior $g(t)$ from the acceptance calculation allows sampling from the distribution even when the maximum lcfit branch length is infinite (i.e., regime 4), since the asymptotic maximum likelihood can still be calculated.

\subsection{Fitting methods}

We have found it useful to use a combination of two methods for fitting.
The first, which we call lcfit4, simply applies standard nonlinear least-squares optimization to find parameters for $f$ using a sample of true values from the original likelihood function.
The second, which we call lcfit2, uses the parameterization in terms of $c$, $m$, $t_0$, and $f''(t_0)$.
In this case we simply set the $t_0$ and $f''(t_0)$ values to their values in the original function, then use least-squares fitting for $c$ and $m$ with a useful set of sampled points (inspired by \cite{Aberer2016-dr}; see below for details).
For lcfit4, we first try unconstrained optimization using the Levenberg-Marquardt (L-M) algorithm \citep{levenberg1944method,marquardt1963algorithm} implemented in the GNU Scientific Library version 1.16 \citep{galassi2003gnu}.
If the L-M algorithm fails to converge to a valid model, we fall back on constrained optimization using the SLSQP algorithm \citep{kraft1994slsqp} implemented in NLopt version 2.4.2 \citep{Johnson2014-ww}.
We have found that trying the L-M algorithm first yields better results in the case of four-parameter optimization than using SLSQP alone.
For lcfit2, we use only the SLSQP algorithm, as we did not find the L-M step necessary to achieve good results.
These two methods are used together as described below.

Next we describe the fitting process for these two methods in more detail.
If one only wants a rough estimate of the likelihood curve, one can simply take a number of pre-chosen points, such as 0.05, 0.1, 0.5, and 1, calculate the corresponding likelihoods, and fit parameters of the curve using least squares as previously described.
On the other hand, if a more accurate likelihood curve is desired, one can use an iterative algorithm to obtain an improved estimate of the likelihood curve around the maximum likelihood branch length.
The idea of this process is to sample until the points enclose the maximum likelihood point.
We will call this method ``lcfit4 fitting''.

First, we fit the initial model:
\begin{enumerate}
  \item Initialize with four values of $t$, and corresponding log likelihoods $\ell$.
  \item If the $\ell$ values are monotonically increasing, add a point: $t = 2 \max(t)$, with corresponding log likelihood.
  \item If the $\ell$ values are monotonically decreasing, add a point: $t = \min(t)/10$ with corresponding log-likelihood.
  \item Repeat until the points enclose a maximum.
\end{enumerate}
lcfit expects a minimum branch length $\mint$ and maximum branch length $\maxt$ to consider.
Some phylogenetic libraries and applications enforce their own values.
When such values are not available, we have found that a small but nonzero value for $\mint$ (such as $10^{-6}$) works well.
For $\maxt$, choose a significantly large value at which the log-likelihood function can be expected to be nearly flat; we used $20$.
Note that excessively large values of $\maxt$ can affect numerical stability.
The current implementation uses 0.1, 0.5, 1.0, and $\maxt$ as the first four starting points for $t$.
$\maxt$ is included in these points to ensure that the fitted model exhibits good asymptotic performance.
The procedure from the previous section is then used to find a starting point of BSM parameters for the optimization algorithm.

We then enter the second phase, which is repeated until the estimate of the ML branch length changes less than some fixed number.
The first step is to find the maximum-likelihood branch length using \eqref{eq:BSMmlt} for the current BSM parameter estimates, and add it to the set of sampled values.
The model is then re-fit using the optimization algorithm.

When the ML branch length is non-zero, we have found this method to be less robust to corner cases than we desired.
Thus we have developed an alternative means of fitting, which we call ``lcfit2 fitting'', that requires finding the ML value and the second derivative.
As described in the main text and derived below, one can re-express the surrogate function in terms of the $c$ and $m$ parameters from before, along with the ML branch length $t_0$ of the surrogate function and its second derivative there.
Then, one can simply set the $t_0$ and $f''(t_0)$ values of the surrogate function equal to the values found on the original function.

The procedure to find $c$ and $m$ for the lcfit2 surrogate after plugging in $t_0$ and $f''(t_0)$ is as follows.
Starting with a default $c$ and $m$,
\begin{enumerate}
\item\label{ITEMlcfit2start} calculate the inflection point $t_i$ for the model.
\item define $\Delta = |t_0 - t_i|$.
\item let our four $t$ values for fitting be $\{t_0 - \Delta, t_0, t_0 + \Delta, \maxt\}$; if either of $t_0 \pm \Delta$ are outside the interval $(\mint, \maxt)$ then replace them with half the distance from $t_0$ to the interval boundary.
\item\label{ITEMlcfit2end} fit $c$ and $m$ using these four points.
\item repeat steps \ref{ITEMlcfit2start}--\ref{ITEMlcfit2end} once more to refine the model.
\end{enumerate}

Our complete fitting routine, using both lcfit2 and lcfit4, is as follows.
First, maximize the original function on the set of non-negative $t$ values.
The maximum is found using Brent's method \citep{brent1973algorithms}.
Next, estimate the first and second derivatives at the maximum using fourth-order finite difference approximations \citep[Table 25.2]{davis1964numerical}.
If the first derivative is nonzero, use lcfit4 fitting, which we have found converges quickly in this case.
If not, then use lcfit2 fitting.
All least-squares fitting is done using the following gradient of $f$:
\begin{align*}
d f/dc & = \log \left(\frac{1}{2} (1+\theta^{-1})\right) \\
d f/dm & = \log \left(\frac{1}{2} (1-\theta^{-1})\right) \\
d f/dr & = (b+t) \frac{m (\theta+1)-c (\theta-1)}{\theta^2-1} \\
d f/db & = r \frac{(m-c) \theta+c+m}{\theta^2-1}
\end{align*}

We have also found it very advantageous to standardize the height of the surrogate function by subtracting the peak of the original function, so that we are fitting a curve that has maximum value zero.
This leaves the asymptote free to vary.

\subsection{Extended methods: benchmarking}

We evaluated the performance of lcfit on both real and simulated data.

We used \texttt{nestly} \citep{mccoy2012nestly} and the Bio++ 2.2.0 suite \citep{dutheil2006biopp,dutheil2008non} of C++ libraries and binaries to perform simulation.
We began by generating random 10-leaf bifurcating trees using the function \texttt{rtree} from the R package \texttt{ape} \citep{paradis2004ape}, with branch lengths sampled from an exponential distribution.
We generated one set of trees with the exponential mean $\mu = 0.1$, and another set with $\mu = 0.01$.
Each set contains 10 independent replicates.
For each tree, we generated a 1000-site sequence alignment with \texttt{bppseqgen} from the Bio++ suite using an evolutionary model from Table~\ref{TABLEmodels} and a rate distribution of either uniform or discretized gamma ($n = 4, \alpha = 0.2$).
We then optimized the branch lengths of each tree with \texttt{bppml}.
The evolutionary model, tree, and alignment were then fed into our \texttt{lcfit-compare} utility.
\texttt{lcfit-compare} loops over each branch of the tree and uses Bio++ to get an empirical log-likelihood function parameterized by the branch length.
It then fits an lcfit model to the empirical log-likelihood function, and both the empirical and surrogate log-likelihood functions are sampled in the neighborhood of the peak.

We estimated the Kullback-Leibler (KL) divergence from the original likelihood function to the surrogate function by sampling these functions over 501 evenly spaced points in the neighborhood of the peak.
This neighborhood is found as the region where the log-likelihood curve is above 10\% of its peak value, bounded by $\mint$ and $\maxt$.
Probabilities are computed from the relative log-likelihoods as
$$
P_i = \frac{\exp(\ell(t_i) - \ell(t_0))}{\sum_j \left[ \exp(\ell(t_j) - \ell(t_0)) \right] }
$$
and
$$
Q_i = \frac{\exp(f(t_i) - f(t_0))}{\sum_j \left[ \exp(f(t_j) - f(t_0)) \right] }
$$
where $t_0$ is the maximum-likelihood branch length.
The KL divergence from the discretized model distribution $Q$ to the discretized empirical distribution $P$ is then calculated as
$$
D_{KL}(P \Vert Q) = \sum_i P_i \log_2 \left( \frac{P_i}{Q_i} \right).
$$

Instructions for running these simulations and the analysis can be found in the \texttt{sims} subdirectory of the lcfit repository at \url{https://github.com/matsengrp/lcfit}.

We also tested the performance of lcfit on real data, in the manner of \cite{Aberer2016-dr}, and compared lcfit to the gamma and Weibull proposal distributions described in their work.
To accomplish this, we incorporated lcfit fitting directly into the ExaBayes code used to generate data for their analysis.
We then compared these results to the ExaBayes results, which were shared with us by Andr\'e Aberer.
Our fork of ExaBayes 1.3.1 used for these experiments can be found at \url{https://github.com/matsengrp/exabayes-1.3.1-lcfit}.
We tested 12 out of the 14 DNA datasets they examined.
One of the datasets not included in our analysis (dat-354) was missing gamma and Weibull distribution fit parameters in the data provided for some edges of the tree.
The other dataset not included (dat-125) yielded a few invalid estimated acceptance rates (i.e., much greater than 100\%).
We attributed these errors to a numerical stability issue in the estimated acceptance rate calculations, and chose to omit the dataset from the analysis entirely rather than present a subset of its results.
The remainder of the datasets contain between 24 and 500 taxa, with sequence lengths ranging from approximately 100 to 30,000 bases.
We reproduced the estimated acceptance rate calculations for gamma and Weibull proposals using the method described in their supplemental material, then applied the same method to lcfit proposals.
We then used the aggregated results to produce Fig.~\ref{FIGacceptance} (analogous to Fig.~2 in \cite{Aberer2016-dr}).

\subsection{Relationship to entropy}

Here we establish a simple relationship between the ML value of the surrogate function and Shannon entropy of a corresponding sequence alignment under the BSM model.
This is not used in practice, but is simply provided here for interest.
Continuing in the setting of the lcfit2 parameterization and with that same notation,
\begin{align*}
\tilde{\theta}^{-1} & = \exp(-r \tilde t)
\end{align*}
such that
$$
f(\tilde{t}) = c \log \left(c+m+ \nu \right) + m \log \left(c+m-\nu \right) - (c+m) \log (2 (c+m))
$$
where
$$
\nu := \frac{c-m}{\tilde{\theta}}. 
$$

At $t = t_0$, $\tilde{\theta} = 1$, so the corresponding $\nu_0 = c - m$.
Also,
\begin{align*}
f(t_0) & = c \log (c + m + \nu_0) + m \log (c + m - \nu_0) - (c + m) \log (2(c + m)) \\
& = c \log (2c) + m \log (2m) - (c + m) \log (2(c + m)) \\
& = c \log c + m \log m - (c + m) \log (c + m).
\end{align*}
Shannon entropy is defined as
$$
S := - \sum_i p_i \log p_i.
$$
Since the $(c + m)$ sites in the model are i.i.d., consider that the probability of observing a substitution at a single site is $p = m / (c + m)$, and the probability of observing no substitution is $1 - p = c / (c + m)$.
Then
\begin{align*}
S & = - \left[ (1 - p) \log (1 - p) + p \log p \right] \\
& = - \left[ \frac{c}{c + m} \log \left( \frac{c}{c + m} \right) + \frac{m}{c + m} \log \left( \frac{m}{c + m} \right) \right] \\
& = - \frac{1}{c + m} \left[ c \log c + m \log m - (c + m) \log (c + m) \right] \\
& = - \frac{1}{c + m} f(t_0).
\end{align*}

\section{Supplementary tables}

\TABLEmodels

\TABLEkloutliers

\notarxiv{\newpage\FIGparams}
\notarxiv{\newpage\FIGsteps}
\notarxiv{\newpage\FIGregimes}

\end{document}